\documentclass[%
 apl,%
 amsmath,amssymb,
 reprint,%
 longbibliography
]{revtex4-1}

\usepackage{graphicx}
\usepackage{dcolumn}
\usepackage{bm}
\usepackage[mathlines]{lineno}
\usepackage{hyperref}
\usepackage{xcolor}
\hypersetup{
    colorlinks=true,
    linkcolor=blue,
    citecolor=blue,
    urlcolor=blue,
}
\usepackage{color,soul}

\usepackage[T1]{fontenc}
\usepackage{siunitx}
\DeclareSIUnit\torr{Torr}
\DeclareSIUnit\oersted{Oe}
\begin{document}


\title[]{Compositional effect on auto-oscillation behavior of Ni$_{\text{100-x}}$Fe$_{\text{x}}$/Pt spin Hall nano-oscillators}
\author{M. Haidar}
\email{\textcolor{black}{corresponding author:} mh280@aub.edu.lb}
 \affiliation{Department of Physics, American University of Beirut, P.O. Box 11-0236, Riad El-Solh, Beirut 1107-2020, Lebanon}
\author{H. Mazraati}
 \affiliation{NanOsc AB, Kista 164 40, Sweden}
 \affiliation{Department of Materials and Nanophysics, School of Engineering Sciences, KTH Royal Institute of Technology, 100 44 Stockholm, Sweden}
\author{P. D\"urrenfeld}%
\affiliation{Department of Physics, University of Gothenburg, 412 96 Gothenburg, Sweden
}%

\author{H. Fulara}%
\affiliation{Department of Physics, University of Gothenburg, 412 96 Gothenburg, Sweden
}%

\author{M. Ranjbar}%
\affiliation{Department of Physics, University of Gothenburg, 412 96 Gothenburg, Sweden
}%

\author{J. \AA{}kerman}
\affiliation{Department of Physics, University of Gothenburg, 412 96 Gothenburg, Sweden
}%
\affiliation{Department of Materials and Nanophysics, School of Engineering Sciences, KTH Royal Institute of Technology, 100 44 Stockholm, Sweden}

\date{\today}
\begin{abstract}

We demonstrate the compositional effect on the magnetodynamic and auto-oscillations properties of Ni$_{\text{100-x}}$Fe$_{\text{x}}$/Pt ($x$ = 10 to 40) nanoconstriction based spin Hall nano-oscillators. Using spin-torque ferromagnetic resonance (ST-FMR) performed on microstrips, we measure a significant reduction in both damping and spin Hall efficiency with increasing Fe content, which lowers the spin pumping contribution. The strong compositional effect on spin Hall efficiency is primarily attributed to the increased saturation magnetization in Fe-rich devices. As a direct consequence, higher current densities are required to drive spin-wave auto-oscillations at higher microwave frequencies in Fe-rich nano-constriction devices. Our results establish the critical role of the compositional effect in engineering the magnetodynamic and auto-oscillation properties of spin Hall devices for microwave and magnonic applications. 





\end{abstract}

\maketitle

The recent demonstration of pure spin current-induced spin-transfer torque arising from the spin Hall effect (SHE) represents an efficient route to controlling the magnetization dynamics in magnetic nanostructures \cite{Hirsh1999,D'yakonov1971}. In a ferromagnet (FM)/heavy metal (HM) bilayer, a pure spin current is generated when a longitudinal charge current passes through the HM and induces a transverse spin current due to the strong spin-orbit coupling in the HM. The conversion of the charge current density ($J_\text{C}$) to pure spin current density ($J_\text{S}$) is characterized by the spin Hall angle ($\theta^{\text{SHA}}$) in these bilayers. The pure spin current may be sufficient to excite perpetual self-oscillations in the magnetization, which can further induce spin waves in these systems. This is of particular interest for spintronics applications, and has led to a new class of microwave devices called spin Hall nano-oscillators (SHNOs) \cite{DemidovSHNO,Ranjbar2014,DemidovNCSHNO,Chen2016,Awad2016,Haidar2019,Fulara2019,Fulara2020,Zahedinejad2020}. Generating higher spin current densities through a higher $\theta^{\text{SHA}}$ is necessary to improve the performance of SHNOs and to avoid higher charge current densities. So far, studies have focused on different HMs with various  spin-orbit coupling strengths to generate higher spin current---for example, the $\beta$-phase of W \cite{Mazraati2016,Demasius2016,Zahedinejad2018apl}, Ta \cite{Liu2012,Dhananjay2017} or Ni$_\text{x}$Cu$_\text{1-x}$ \cite{Kelller2019}. It has  also been shown that modifying the interface with Hafnium \cite{Pai2014,Mazraati2018} or an antiferromagnet \cite{WangAFM,Hahn2014} enhances the transmission of the spin current between the HM and the FM. The amplitude of the spin current generated in the FM/HM bilayer is also controlled by the magnetodynamics of the ferromagnet \cite{Yoshino2011,Yuli2015} and the transparency of the FM--HM interface \cite{Zhang2015}. 

In this work, we use spin-torque-induced ferromagnetic resonance (ST-FMR) and auto-oscillation measurements to study the impact of  alloy composition on the magnetodynamics and  spin-torque properties of spin Hall devices made of Ni$_{\text{100-x}}$Fe$_{\text{x}}$~(5~nm)/ Pt~(6~nm) bilayers, where x~=~[10--40] denotes the Fe content in the FM in percent. First, we measure an  increase in  the magnetization, a decrease in the Gilbert damping, and a reduction of the impact of spin-torque on the ST-FMR linewidth as Fe content increases. Interestingly, we find a reduction of the spin Hall angle in Fe-rich devices that scales in inverse proportion to the magnetization. Then, we study the microwave auto-oscillation characteristics as a function of Fe-content in nanoconstriction SHNOs. In Fe-rich devices, the auto-oscillations have higher frequencies accompanied with low output power, in addition, higher threshold current densities are required to drive auto-oscillations.   

\begin{figure}[!h]
\begin{center}
\includegraphics[width=0.48\textwidth]{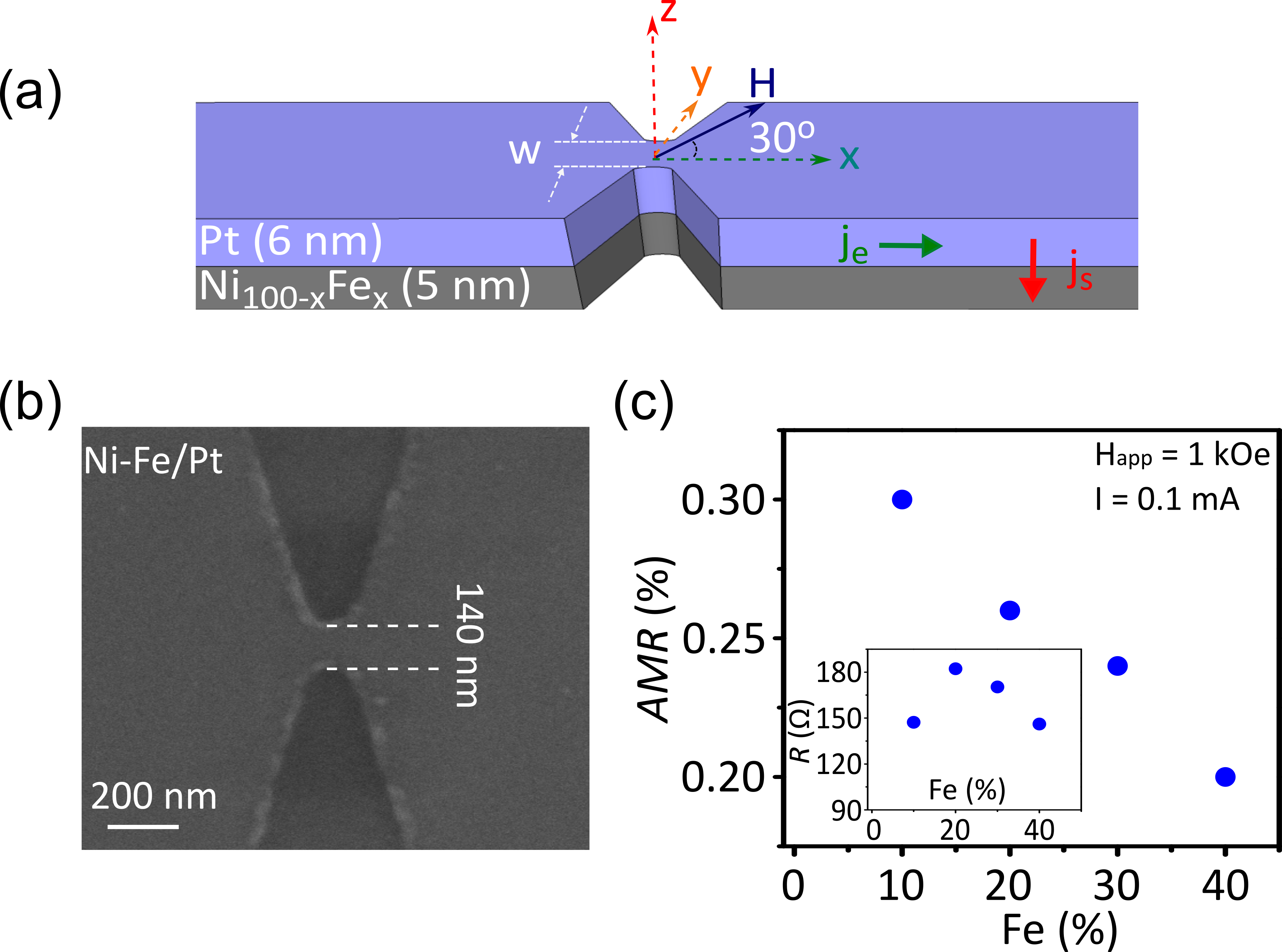}
\caption{\label{fig1} (a) Schematic layout of  nanoconstriction devices made of  Ni$_{\text{100-x}}$Fe$_{\text{x}}$/Pt bilayers. This shows the direction of the applied field and the dc current. (b) Scanning electron microscope image of a device with 140 nm width. (c) Anisotropic magnetoresistance and resistance (inset) of the devices as a function of Fe content.}.
\end{center}
\end{figure}

The layout of our devices is sketched in Figure~\ref{fig1}(a). The devices are fabricated from 
Ni$_{\text{100-x}}$Fe$_{\text{x}}$/Pt bilayers deposited in a high vacuum magnetron sputtering chamber with a base pressure below \SI{2e-8}{\torr} on sapphire C-plane substrates at room temperature. \textcolor{black}{The different Ni-Fe alloys were co-sputtered under the same conditions and from the same pure Ni and Fe targets, where the composition was established by varying the respective plasma powers.} 
Energy-dispersive x-ray spectroscopy (EDX) measurements showed that the actual alloy composition is within $\pm$ 3 atomic percent error of the nominal composition. Two kinds of spin Hall devices were fabricated from each film: (1) \SI[product-units = power]{8 x 16}{\um} rectangular stripes and (2) nanoconstriction-based SHNOs. The nanoconstriction devices are designed as \SI[product-units = power]{4 x 12}{\um} rectangles with two indentations of \SI{50}{\nm} tip radius forming a nanoconstriction in the center of the strip with a nominal width of \SI{140}{\nm}. For Ni$_{\text{60}}$Fe$_{\text{40}}$/Pt device, the nominal width of the nano-constriction is \SI{120}{\nm}. These patterns were fabricated using electron beam lithography followed by an etching process. The devices are protected from long-term degradation by a \SI{50}{\nm} thick SiOx layer. A scanning electron micrograph of the nanoconstriction is shown in Figure~\ref{fig1}(b). We established electrical connection to the devices using coplanar waveguides, which we fabricated through a lift-off process from \SI{1}{\um} copper and \SI{20}{\nm} gold. Figure~\ref{fig1}(c) shows the anisotropic magnetoresistance (AMR) and the resistance (inset) of the nanoconstriction devices as a function of the percentage of Fe. We measured a reduction in  AMR in Fe-rich devices.

\begin{figure}[!h]
\begin{center}
\includegraphics[width=8.5 cm]{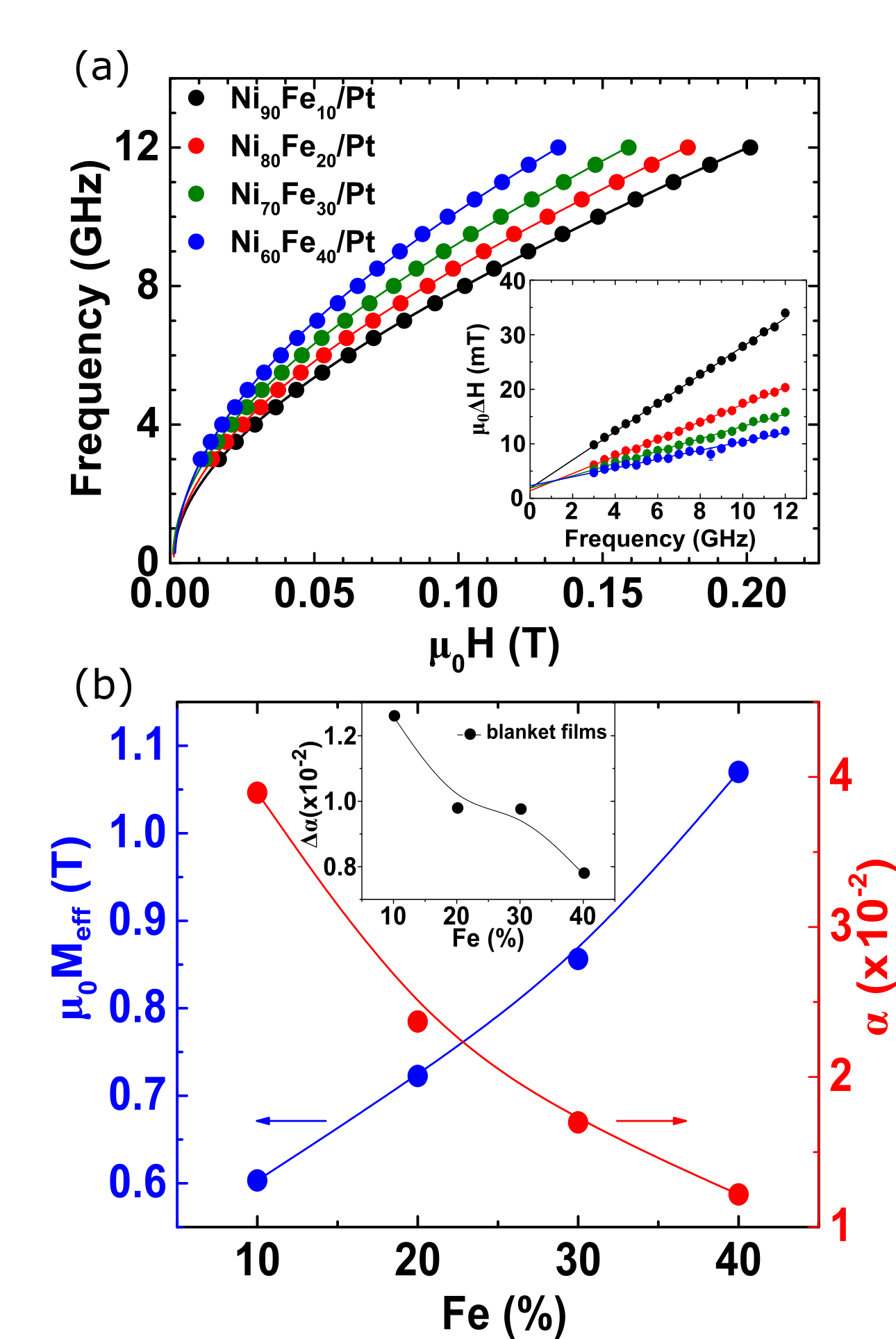}
\caption{\label{fig2} Magnetodynamics of rectangular stripes.(a) Variation of  frequency vs. field and linewidth vs. frequency (inset) for Ni$_{\text{100-x}}$Fe$_{\text{x}}$/Pt devices. (b) shows the magnetization (blue circles) and the Gilbert damping (red circles) as a function of the Fe content. Inset, variation of $\Delta \alpha$ vs. Fe content, the difference in damping of Ni$_{\text{100-x}}$Fe$_{\text{x}}$ films with and without the Pt layer, measured on similar blanket films using broadband-FMR. 
}
\end{center}
\end{figure}

\begin{table*}[!ht]
\centering
 \resizebox{0.65\textwidth}{!}{
\begin{tabular}{ c | c| c | c | c | c | c }
\hline
Samples & $\mu_{0}M_{s}$& $\gamma/2\pi$& $\mu_{0}M_{\text{eff}}$& $\mu_{0}H_{k}$ & $\alpha$ & $R_{FM}$\\
&(T) &(\si{\GHz/T})& (T) &(mT)&($10^{-2}$)& ($\Omega$)\\
\hline
Ni$_{\text{90}}$Fe$_{\text{10}}$/Pt & 0.67 &30& 0.673 & 1.1 & 3.89 $\pm$ 0.1& 98.5\\
\hline
Ni$_{\text{80}}$Fe$_{\text{20}}$/Pt &0.82 & 30&   0.819 & 1.2 & 2.37 $\pm$ 0.1& 185.6  \\
\hline
Ni$_{\text{70}}$Fe$_{\text{30}}$/Pt &0.926 &30&   0.926 & 0.8 & 1.69 $\pm$ 0.1& 152.2 \\
\hline
Ni$_{\text{60}}$Fe$_{\text{40}}$/Pt &1.18 &30&   1.074  & 1.7 & 1.2 $\pm$ 0.1& 93.3
\\
\hline
\end{tabular}}
\caption{\label{table1}Summary of the magnetodynamics for Ni$_{\text{100-x}}$Fe$_{\text{x}}$/Pt rectangular stripe devices.}
\end{table*}

We first discuss the ST-FMR spectra measured on rectangular stripes to determine the variation of the magnetodynamics with the Fe content. The ST-FMR measurements were performed by connecting the devices to a pulse-modulated signal generator to excite the dynamics, and to a lock-in amplifier to detect the mixing voltage, which arises in resonance due to the synchronism of the time-varying magnetoresistance with the excitation current. The devices are placed in a uniform magnetic field applied in-plane at an angle of \ang{30} with respect to the x-direction. The measurements were performed by sweeping the magnetic field at a fixed frequency to reduce the nonmagnetic background. The measurements were performed over a broad range of frequencies.

We observe that the magnitude of the dc voltage around the resonance field depends on the Fe percentage composition. The shape of the measured dc voltage originates from two processes that can be distinguished by their symmetry with respect to the field. The asymmetric part originates from the anisotropic magnetoresistance (AMR), while the  symmetric part comes from the spin torque and  inverse spin Hall effect voltage.\cite{Liu2011} We thus fit the dc voltage to a single Lorentzian function which has symmetric and antisymmetric components:
\begin{equation}
V = V_0+ V_{\text{S}}\frac{\Delta H^{2}}{\Delta H^{2}+4(H-H_{\text{r}})^2}+V_{\text{A}} \frac{4\Delta H (H-H_{\text{r}})}{\Delta H^{2}+4(H-H_{\text{r}})^2},
\label{equation1}
\end{equation} 
where $V_0$ is the  background voltage, $V_{\text{S}}$ and $V_{\text{A}}$ are the the symmetric and antisymmetric Lorentzian coefficients respectively, $H_{\text{r}}$ is the resonance field, and $\Delta H$ is the full width at half maximum. From the fitting, we extract $H_{\text{r}}$ and $\Delta H$ at each frequency $\it{f}$, as plotted in Figure~\ref{fig2}(a), to extract the magnetodynamics of these devices. By fitting $\it{f}$ vs. $H_{\text{r}}$ to the in-plane Kittel equation $f=\frac{\mu_0 \gamma}{2\pi}\sqrt{(H+H_k)(H+H_k+M_{\text{eff}})}$, where $\gamma/2\pi$ is the gyromagnetic ratio and $H_k$ is the in-plane anisotropy, the effective magnetization $(\mu_0M_{\text{eff}})$ of device can be extracted. A clear reduction in $\mu_0 M_{\text{eff}}$ is measured for Ni-rich films, as shown in Figure~\ref{fig2}(b), as the magnetic moment per atom decreases in Ni-rich films. We found only small values of $\mu_{0}H_k$ (Table~\ref{table1}). We measure the saturation magnetization $(\mu_0 M_{s})$ separately on blanket films, using Alternating Gradient Magnetometry (AGM) (Table~\ref{table1}).

The Gilbert damping ($\alpha$) can be extracted from the slope of $\mu_0 \Delta H$ vs. $f$ using $\mu_0 \Delta H= \frac{2 \alpha}{\gamma} (2 \pi \it{f}) +\Delta H_{0}$, where $\Delta H_{0}$ is the inhomogeneous broadening. The results for the $\alpha$ vs. Fe composition are shown in Figure~\ref{fig2}(b). Several mechanisms determine the measured value of $\alpha$. An intrinsic contribution with a similar increase in the Gilbert damping has been reported in bulk Ni-Fe alloys.\cite{Bonin2005,Starikov2011} In Ni-rich films, the  larger values of $\alpha$ can be understood in terms of a conductive-like behavior  that can be explained by intraband scattering.\cite{Gilmore2007} 
In addition to the intrinsic origin of $\alpha$ in our devices, the spin pumping contributes substantially to the measured damping of each device due to the strong spin-orbit coupling of Pt. 
We compare the damping of Ni$_{\text{100-x}}$Fe$_{\text{x}}$ with and without the Pt layer on a similar blanket films, where we estimated the spin pumping contribution to the damping $\Delta \alpha =(\alpha_\text{{FM/Pt}}-\alpha_\text{{FM/Cu}})$. We measure a decrease in $\Delta \alpha$ with the Fe-content as shown in the inset of inset of Figure~\ref{fig2}(b), which suggests that bilayers containing Fe-rich films have lower spin pumping.

\begin{figure}\centering
\includegraphics[width=8.5 cm]{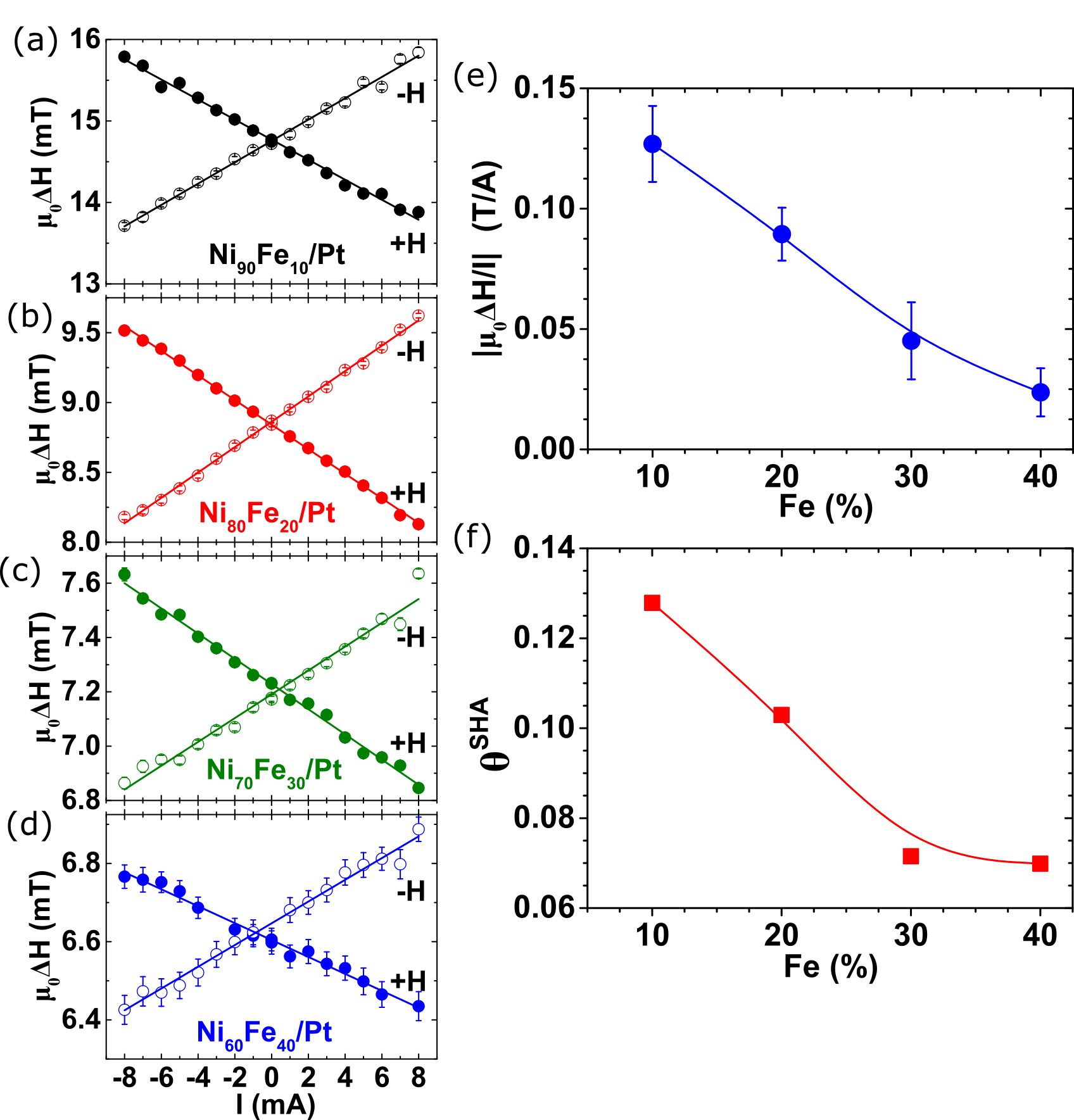}
\caption{\label{fig3} (a--d) The current induces a change in  the linewidth $\mu_0\Delta H$ as a function of the dc current for different Ni-Fe compositions. Closed and open circles show results of measurements performed  under  +H and -H respectively. (e--f) Variation in $|\mu_0\Delta H/I|$ and the spin Hall angle ($\theta^\text{{SHA}}$) as a function of Fe composition. 
}
\end{figure}

Next, we  discuss the correlation between the spin Hall efficiency and the Fe composition. We carried out ST-FMR measurements on rectangular stripes spin Hall devices at $f$ = 5~GHz with an injected power of $P$ = \mbox{+15}~dBm. We examined the variation of $\mu_0 \Delta H$ with an applied dc current ($I$) in the range of +8 mA to -8 mA for all devices. The ST-FMR signal reverses its sign, as we sweep the direction of the field  or the current direction to the opposite polarity. 
Figure~\ref{fig3}(a--d) shows the variation of the extracted $\mu_0\Delta H $ vs. $I$ for devices measured at +$H$ (closed circles) and -$H$ (open circles). A behavior common  to these measurements is that, for a positive field, $\mu_0 \Delta H$ decreases with +$I$ and increases with -$I$. We observe the opposite behavior under negative fields, as expected from the symmetry of the spin Hall effect. This is due to the spin-torque exerted by the pure spin currents generated from the SHE in Pt and injected into the FM layer. The spin-torque is aligned either parallel or antiparallel to the Gilbert damping, which leads to an effective enhancement or reduction in the damping of the ferromagnetic layer. The linewidth analysis is a reliable method for measuring the spin Hall angle, as it originates only from the antidamping spin-transfer torque. In the linear regime, the current-induced spin torque could be modeled as:\cite{Liu2011,haidar2016}
\begin{equation}
\begin{array}{rcl}
\mu_0 \Delta H &=& \frac{2\pi f}{\gamma} 2\alpha +\Delta H_{0} \\
& & +\frac{2\pi f}{\gamma \left( H+2\pi M_{\text{eff}} \right)}
\frac{\sin \left( \varphi \right)}{\mu_{0}M_{s}t} \frac{\hbar }{2 e A_c}\frac{R_{FM}}{R_{FM}+R_{Pt}}\theta^{SHA} I, 
\label{LW}
\end{array}
\end{equation}
where $R_{FM}$ and $R_{Pt}$ are the resistance of the FM layer and the Pt, respectively. $A_c$ is the cross section of the device, $\varphi$ is the in-plane angle, and $\theta^\text{{SHA}}$ is the spin Hall angle of the FM/Pt bilayer that includes possible angular momentum losses at the interfaces. The first and the second terms represent the linewidth in the absence of the dc current $\mu_0\Delta H (I =0)$, while the third term on the right-hand side is the spin current-induced modification to $\mu_0\Delta H$. 

The solid lines in Figure~\ref{fig3}(a--d) are linear fits to Equation~(\ref{LW}), from which we extracted the slope $|\mu_0 \Delta H /I|$. Figure~\ref{fig3}(e) summarizes the results of the extracted $|\mu_0 \Delta H /I|$ for each device as a function of Fe composition. The impact of spin-torque on the ST-FMR linewidth is found to be linear with a slope $|\mu_0 \Delta H /I|$ that decreases by a factor of six with the Fe content. We then calculated the $\theta^\text{{SHA}}$ from $|\mu_0 \Delta H /I|$ using Equation~(\ref{LW}), since the other factors are constants in our measurements. We measure $R_{Pt}$ = 65 $\Omega$ and $R_{FM}$ values are summarized in (Table~\ref{table1}). Figure~\ref{fig3}(f) shows $\theta^\text{{SHA}}$ as a function of Fe content where a reduction in $\theta^\text{{SHA}}$ by a factor of two is measured for Fe-rich devices.
$\theta^\text{{SHA}}$ of Ni$_{80}$Fe$_{20}$/Pt is similar to the reported values of permalloy-based bilayers. \cite{Liu2011}
As the magnetization increases from 0.67 T for Ni$_{\text{90}}$Fe$_{\text{10}}$/Pt to 1.18 T for Ni$_{\text{60}}$Fe$_{\text{40}}$/Pt, roughly by a factor of two, the $\theta^\text{{SHA}}$ decreases by a factor two. 
This hints that the reduction in the spin Hall angle can be correlated primarily to the compositional effect: the spin Hall angle scales inversely proportional with the saturation magnetization. 
Other sources may contribute to the reduction of $\theta^\text{{SHA}}$: \emph{i}) the transparency of the Pt--ferromagnet interface to the spin current. \cite{Zhang2015, Caminale2016} \emph{ii}) Part of the spin Hall efficiency may also originate from the Ni$_{\text{1-x}}$Fe$_{\text{x}}$ layer itself,\cite{haidar2013,Haidar2019} as the spin Hall angles of Ni and Fe have opposite signs.\cite{Wang2014}

\begin{figure}[!h]
\begin{center}
\includegraphics[width=8.5 cm]{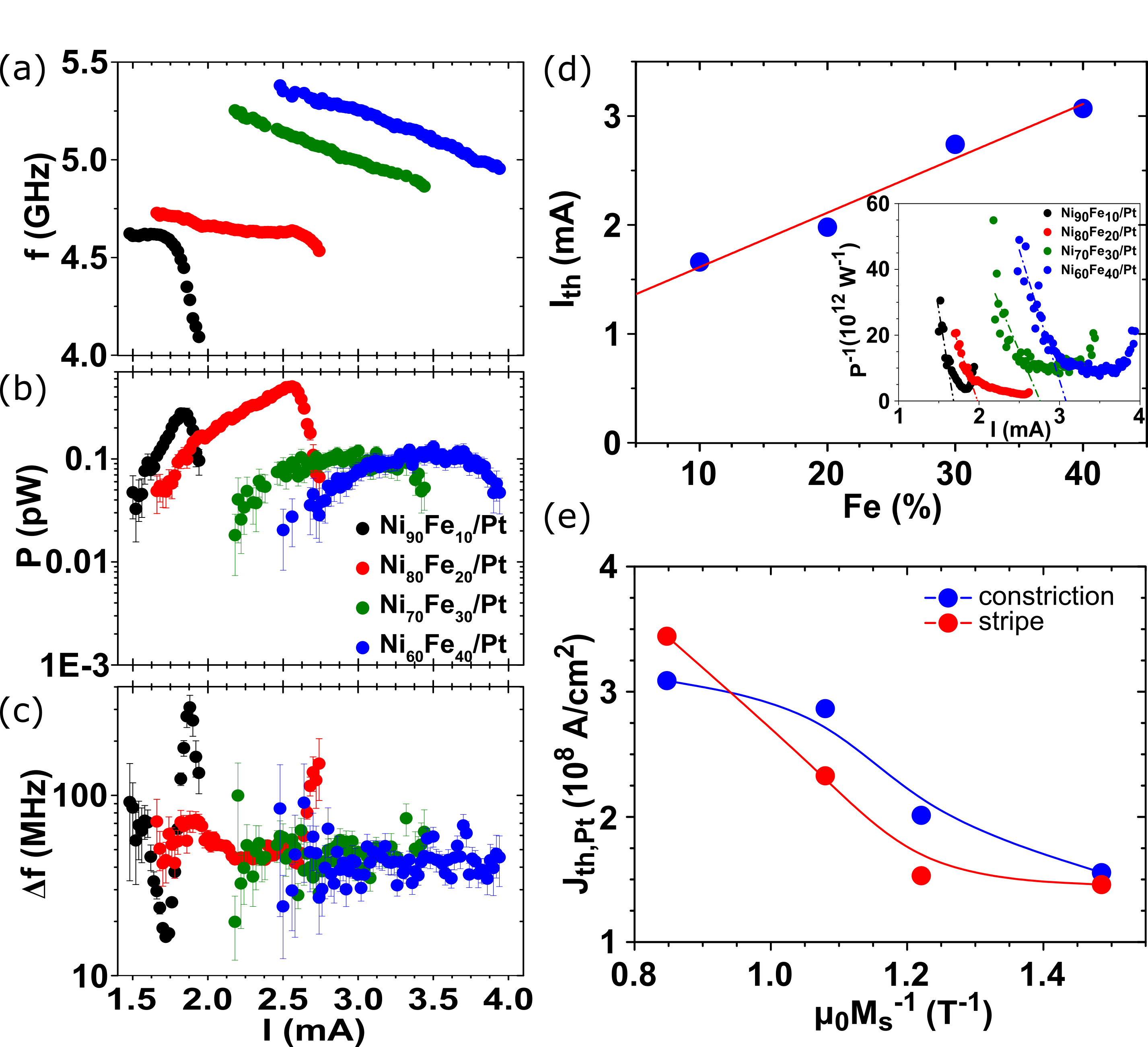} 
\caption{\label{fig4} (a) Frequency, (b) integrated power, and (c) linewidth of the microwave auto-oscillations as a function of current for different SHNOs. The applied field is $\mu_{0}H_{ext}$ = 0.05 T. (d) Threshold current vs. Fe concentration. Inset shows how the threshold is extracted from a linear fit of $p^{-1}$ vs.~$I$ at low currents (dashed line). (e) Threshold current density in Pt vs. $(\mu_0M)^{-1}$ extracted from ST-FMR (red circles) and auto-oscillation (blue circles) measurements. 
}
\end{center}
\end{figure}

Next, we turn to discuss the compositional effect on the characteristics of auto-oscillations in nanoconstriction-based SHNOs. To study auto-oscillations, we replace the lock-in amplifier with a spectrum analyzer and a low noise amplifier with a +33 dB gain. The generated microwave power spectral density (PSD) was recorded as a function of a direct current under an in-plane magnetic field $\mu_0 H = 0.05$ T. The auto-oscillation frequencies, integrated power and linewidth are extracted by fitting the peaks of all spectra with a Lorentzian function. Figure~\ref{fig4}(a-c) shows the spectral characteristics of the microwave auto-oscillations. As the device is biased with a direct current, the emission of auto-oscillation starts once the spin-torque overcomes the damping completely. We observe emission of auto-oscillation at a single frequency, below FMR, at the threshold current and then at higher currents a red-shift in frequencies is detected. 
It is interesting to note that the onset auto-oscillation frequency differs between devices, i.e. the onset frequency shifts up for Fe-rich devices due to higher magnetization. The integrated power varies with the dc current, showing a bell shape of an amplitude around 1 pico-watt. A qualitative comparison between the power of devices shows that Fe-rich devices have lower output power. This is understood as the readout of devices depends on AMR and as the AMR drops in Fe-rich devices the power follows. Finally, the measured auto-oscillations have linewidths around 50 MHz. 
The  threshold  current for auto-oscillations, $I_{th}$, is extracted  from plots of $p^{-1}$ vs.~$I$, \cite{slavin2009} as shown in the inset of Figure~\ref{fig4}(d). The dashed lines show the position of the $I_{th}$, obviously, for Fe-rich devices higher currents are required to excite auto-oscillations as shown in Figure~\ref{fig4}(d). From $I_{th}$, we estimate the threshold current density in Pt, $J_{th,Pt}$ , as plotted in Figure~\ref{fig4}(e) (blue dots). It is also interesting to compare $J_{th,Pt}$ with that extracted from the ST-FMR measurements on stripes by extrapolating the $\mu_0\Delta H$ vs. I in Figure~\ref{fig3}(e), to zero. Results are shown in Figure~\ref{fig4}(e) (red dots). A perfect matching is measured between $J_{th,Pt}$ extracted from both measurements. We observe an increase in $J_{th,Pt}$ with the increase of the magnetization. The enhancement in threshold current densities for Fe-rich devices is a direct consequence of the reduction in the spin Hall angle. Our experimental results show that the dominant compositional effect is from the magnetization that plays a central role in determining the characteristics of the spin Hall devices.

In conclusion, we have investigated the impact of alloy composition on the magnetodynamic properties of Ni-Fe-based spin Hall devices. Our results will be useful not only for fundamental studies of spin transport between Ni-Fe alloys and heavy metal layers, but also in the engineering of new devices with controllable properties for spin torque and magnonic devices. 

This work was supported by the Swedish Research Council (VR; Dnr.~2016-05980), the Knut and Alice Wallenberg foundation (KAW), and  the  American  University  of  Beirut Research Board (URB). This work was also partially supported by the European Research Council (ERC) under the European Community's Seventh Framework Programme (FP/2007-2013)/ERC Grant 307144 "MUSTANG".

The data that support the findings of this study are available from the corresponding author upon reasonable request.

\bibliographystyle{aipnum4-1}
\bibliography{biblioth}

\end{document}